\documentclass[nofootinbib,aps,a4paper,letterpaper,superscriptaddress,
twocolumn,showpacs,times,eqsecnum]{revtex4}
\usepackage{amsmath}
\usepackage{amsfonts}
\usepackage{booktabs}
\usepackage{multirow}
\usepackage{siunitx} 
\usepackage{adjustbox}
 \pdfoutput=1
\usepackage{graphicx}
\usepackage{color}
\usepackage{braket}
\usepackage{dcolumn}
\usepackage{bm,url}
\usepackage[linktocpage]{hyperref}
\usepackage{subfigure}
\usepackage{amsfonts}
\usepackage[usenames,dvipsnames,svgnames]{xcolor}  
\usepackage{hyperref}   
\definecolor{oxfordblue}{rgb}{0.0, 0.13, 0.28}
\definecolor{burgundy}{rgb}{0.5, 0.0, 0.13}
\definecolor{darkolivegreen}{rgb}{0.33, 0.42, 0.18}
\definecolor{darkblue}{rgb}{0,0,0.5}
\definecolor{richcarmine}{rgb}{0.84, 0.0, 0.25}
\definecolor{darkblue}{rgb}{0,0,0.5}
\definecolor{bluer}{rgb}{0.00,0.50,0.75}{}
\hypersetup{colorlinks=true, citecolor=red, linkcolor=blue,
 urlcolor = magenta, filecolor=magenta}

\begin{document}
 
 \newcommand\be{\begin{equation}}
  \newcommand\ee{\end{equation}}
 \newcommand\bea{\begin{eqnarray}}
  \newcommand\eea{\end{eqnarray}}
 \newcommand\bseq{\begin{subequations}} 
  \newcommand\eseq{\end{subequations}}
 \newcommand\bcas{\begin{cases}}
  \newcommand\ecas{\end{cases}}
 \newcommand{\p}{\partial}
 \newcommand{\f}{\frac}

 \title{Cosmologically viable non-polynomial quasi-topological gravity: 
explicit models, $\Lambda$CDM limit and observational constraints}

 \author{Emmanuel N. Saridakis}
 \email{msaridak@noa.gr}
 \affiliation{Institute for Astronomy, Astrophysics, Space Applications and 
Remote Sensing, National Observatory of Athens, 15236 Penteli, Greece}
 \affiliation{Departamento de Matem\'{a}ticas, Universidad Cat\'{o}lica del 
  Norte, Avda. Angamos 0610, Casilla 1280, Antofagasta, Chile}
 \affiliation{CAS Key Laboratory for Research in Galaxies and Cosmology, 
School 
  of Astronomy and Space Science,
  University of Science and Technology of China, Hefei 230026, China}

\begin{abstract}
We investigate the cosmological implications of  a phenomenologically
relevant subclass  of non-polynomial quasi-topological gravity (NPQTG),
in which the background dynamics is encoded in a single function of the
Hubble parameter.  This framework provides a 
minimal and theoretically consistent extension of general relativity, 
incorporating higher-curvature effects while preserving second-order 
field equations and avoiding higher-derivative instabilities.
We first establish the general conditions for cosmological viability and 
construct explicit realizations, including polynomial, quartic, 
power-law and non-polynomial models, demonstrating how different functional 
forms lead to 
distinct expansion histories. 
Focusing on the quartic and power-law cases, we investigate their
late-time cosmological evolution, showing that they exhibit a
matter-dominated epoch followed by a dark-energy-dominated epoch, and
give rise to an effective dark-energy sector of geometric origin, with
dynamical equation-of-state behavior that can lie in the quintessence or
phantom regime. 
We then confront the models with observational data from Type Ia 
Supernovae, Cosmic Chronometers, and Baryon Acoustic Oscillations, using 
a Bayesian MCMC analysis.
We find that both models are fully compatible with the considered data and
reproduce the late-time expansion history with a quality of fit comparable to
that of $\Lambda$CDM, although the current observations do not provide
statistically significant evidence for a departure from the concordance
scenario.
Our results demonstrate that NPQTG offers a simple and efficient 
framework for describing late-time cosmic acceleration with dynamical 
dark energy, while maintaining theoretical consistency and observational 
viability.
\end{abstract}
 
 \maketitle

\section{Introduction}

General relativity has been proven to be an extremely successful  theory in 
describing the gravitational interaction and the cosmological evolution. 
Nevertheless, theoretical issues, such as non-renormalizability, and 
cosmological progress, such as the discovery of the late-time Universe 
acceleration ~\cite{SupernovaSearchTeam:1998fmf,
SupernovaCosmologyProject:1998vns}   and possible  observational 
tensions \cite{CosmoVerseNetwork:2025alb}, have led to a systematic research 
toward gravitational modifications 
\cite{CANTATA:2021asi,Clifton:2011jh,Capozziello:2011et,Nojiri:2017ncd}. The 
 simplest way to construct  such theories   is by extending the standard 
Einstein-Hilbert action in various ways, obtaining $f(R)$ gravity
 \cite{Starobinsky:1980te, Capozziello:2002rd, DeFelice:2010aj}, 
 $f(G)$ gravity \cite{Nojiri:2005jg, DeFelice:2008wz}, cubic gravity 
 \cite{Asimakis:2022mbe}, Lovelock gravity \cite{Lovelock:1971yv, 
  Deruelle:1989fj}. Alternatively, one can be based on the equivalent torsional 
  formulation of gravity,  and modify it   
 similarly, resulting to $f(T)$ gravity \cite{Cai:2015emx, Ferraro:2006jd,
  Linder:2010py, Chen:2010va,Tamanini:2012hg},  $f(T, T_G)$ gravity 
 \cite{Kofinas:2014owa, Kofinas:2014daa},  $f(T, B)$ gravity
 \cite{Bahamonde:2015zma, Bahamonde:2016grb}, etc, or on the equivalent 
non-metricity framework, resulting to $f(Q)$ theories 
\cite{BeltranJimenez:2017tkd,BeltranJimenez:2019tme,Anagnostopoulos:2021ydo,
Heisenberg:2023lru} and $f(Q,C)$ gravity 
\cite{De:2023xua}.

 On the other hand, a large class of gravitational modifications arises in the 
framework of scalar-tensor theories, where extra scalar fields are introduced 
and are coupled through various ways with geometrical terms, such as in 
Horndeski gravity 
 \cite{Horndeski:1974wa}, in generalized 
Galileon theory \cite{DeFelice:2010nf}, in bi-scalar gravity 
\cite{Saridakis:2016ahq} etc. Furthermore, similar constructions can be 
established in the alternative geometrical formulations of gravity too, 
resulting to 
 scalar-torsion theories 
 \cite{Geng:2011aj}, Teleparallel Equivalent of Horndeski theories   
 \cite{Bahamonde:2019shr, 
  Bahamonde:2020cfv},  scalar-nonmetricity theories 
\cite{Runkla:2018xrv,Bahamonde:2022cmz} etc.

Although the above theories  lead to a rich phenomenology capable of 
addressing both early- and late-time cosmic 
acceleration, such extensions are often accompanied by significant 
theoretical challenges, including the appearance of higher-derivative equations 
of motion, the propagation of ghost-like degrees of freedom, and the loss of 
predictivity due to the proliferation of arbitrary terms.

To overcome these difficulties, the class of quasi-topological gravities was 
introduced in the literature
\cite{Oliva:2010eb,Myers:2010ru}. These theories are 
constructed such that, despite involving higher-curvature terms, they lead to 
second-order equations of motion for highly symmetric spacetimes, including 
static and cosmological backgrounds. In this way, they preserve 
one of the key advantages of general relativity, namely  the absence of 
higher-derivative 
instabilities, while allowing for non-trivial modifications of 
gravitational dynamics 
\cite{Oliva:2010eb,Myers:2010ru,Myers:2010jv,Kuang:2010jc, 
Kuang:2011dy,Dehghani:2013ldu,Chernicoff:2016qrc,Cisterna:2017umf,
Hennigar:2017ego,Dykaar:2017mba,Arciniega:2018tnn,Mir:2019ecg,Mir:2019rik,
Frolov:2024hhe,
Sotkov:2012kx,Ghanaatian:2013wva,Sheykhi:2014rka,Ghanaatian:2016cmf,Lan:2017xcl,
Ghanaatian:2018gdl,Ghanaatian:2018cmz,Fierro:2020wps,Bazrafshan:2020fvn,
Ali:2022oqr,Sekhmani:2022lws,Marks:2023ipa,Olamaei:2023csx,Ali:2023gny,
Bueno:2024dgm,
Ali:2024rtb,Ling:2025ncw,Fan:2025jow,Li:2026mam,Tsuda:2026xjc,Konoplya:2026gim,
Bueno:2026dln,Lutfuoglu:2026gis,Dubinsky:2026wcv,Malik:2026laq,
PinedoSoto:2026hfm,Sueto:2026epz}. 
Nevertheless, in four spacetime dimensions, the simplest versions of such 
constructions, i.e. the polynomial 
quasi-topological gravities, are severely constrained or trivial, limiting 
their applicability in realistic cosmological 
settings \cite{Bueno:2019ltp,Moreno:2023rfl}.

This limitation has recently been overcome through the construction of 
non-polynomial quasi-topological gravities (NPQTG) 
\cite{Colleaux:2015yta,Colleaux:2019ckh,Colleaux:2020wfv,Bueno:2025zaj} (see 
also 
\cite{Chinaglia:2017wim}). These constructions  
arise from generally covariant higher-dimensional gravitational theories and 
admit a consistent dimensional reduction to an effective two-dimensional 
scalar-tensor theory of Horndeski 
type \cite{Borissova:2026krh,Borissova:2026wmn,Borissova:2026klg}. 
Remarkably, for an important subclass of these theories the cosmological
dynamics simplifies drastically, and the full set of field equations can
be reduced to an algebraic relation between a scalar curvature invariant
and the matter energy density. As a result, within this subclass the
expansion history of the Universe is governed by a single function of
the Hubble parameter, incorporating the underlying gravitational
modifications in a compact and tractable form.

The structure of NPQTG leads to significant advantages. Firstly, the 
resulting cosmological equations remain of second order, ensuring theoretical 
consistency and stability.
Secondly, within the subclass considered here, the complexity of 
higher-curvature gravity is effectively resummed into a single function, 
allowing for an efficient parametrization of deviations from general 
relativity.
Thirdly, the framework naturally accommodates non-trivial high-energy 
modifications, making it particularly suitable for addressing early-universe 
phenomena such as inflation, 
bouncing cosmologies, and the resolution of 
singularities  \cite{Borissova:2026klg}, while at the same time 
allowing for a smooth recovery of standard cosmology at late times. In this 
sense, NPQTG provides a minimal and controlled way to probe deviations from 
general relativity 
across different cosmological regimes.

Despite these appealing features, a systematic investigation of the 
cosmological implications and observational viability of NPQTG remains largely 
unexplored. In particular, it is crucial to identify the conditions under which 
these theories reproduce the standard cosmological sequence, admit a consistent 
late-time acceleration phase, and remain compatible with current observational 
data. Furthermore, explicit realizations connecting the underlying 
gravitational action to phenomenologically viable cosmological scenarios are 
still needed.

In the present work, we address these issues by performing a detailed 
cosmological analysis of non-polynomial quasi-topological gravities. We first 
present the theoretical framework and derive the modified Friedmann equations 
governing the cosmological dynamics. We then construct explicit models, both 
polynomial and non-polynomial, illustrating how different functional forms of 
the original theory translate into distinct cosmological behaviors. We 
investigate their cosmological evolution and we  confront 
them with observational data, including Cosmic Chronometers, Type Ia 
Supernovae, and Baryon Acoustic Oscillations, in order to assess their 
viability and constrain the corresponding parameter space.

 \section{Non-polynomial quasi-topological gravity and cosmology}
 \label{NPQTGtheory}

In this section we present the theoretical framework of non-polynomial 
quasi-topological gravity (NPQTG), focusing on its cosmological formulation. 
Our goal is to provide a self-contained description that leads directly from 
the underlying action to the effective Friedmann equation governing the 
background evolution.

\subsection{From higher-dimensional gravity to effective cosmological dynamics}

Constructing modified theories of gravity that consistently depart from general 
relativity at high curvature, while remaining theoretically well-controlled, 
is a central challenge in modern cosmology. In particular, higher-order
extensions are expected to play a crucial role in addressing fundamental 
issues such as the ultraviolet completion of 
gravitational dynamics, or the detailed observational confrontation. However, 
generic higher-order theories typically lead 
to higher-derivative equations of motion and the propagation of additional 
degrees of freedom, often associated with ghost-like instabilities.

Non-polynomial quasi-topological gravity provides a framework in which these 
difficulties can be circumvented in a non-trivial way. The essential feature of 
this construction is that, although it originates from higher-curvature 
invariants, its dynamics in highly symmetric spacetimes, and in particular 
in homogeneous and isotropic cosmologies, reduces to an effectively 
second-order system. As a result, one retains the advantages of 
higher-curvature modifications, while preserving a tractable and physically 
transparent description of the cosmological evolution.

A natural way to realize such theories is through the dimensional reduction of 
generally covariant gravitational actions defined in $d \geq 4$ spacetime 
dimensions \cite{Borissova:2026klg}. In this approach, one considers 
warped-product geometries of the 
form
\begin{equation}
g_{\mu\nu}(x) 
d x^\mu d x^\nu = q_{ab}(y)d y^a d y^b + \varphi(y)^2 d 
\Sigma_{d-2}^2 \,,
\end{equation}
where $q_{ab}(y)$ is a two-dimensional metric and $\varphi(y)$ is a scalar 
field encoding the size of the transverse space of constant curvature.

It has been shown that a broad class of higher-dimensional gravitational 
theories, including quasi-topological ones, reduce consistently to 
two-dimensional scalar-tensor theories of Horndeski type 
\cite{Colleaux:2017ibe,Borissova:2026krh}. This implies that 
solving the reduced system is equivalent to solving the full 
higher-dimensional gravitational theory, while significantly simplifying the 
analysis and making the relevant degrees of freedom manifest.

For cosmological applications, we focus on spatially homogeneous and isotropic 
backgrounds described by the Friedmann-Lemaître-Robertson-Walker (FLRW) metric
\begin{equation}
d s^2 = -d t^2 + a(t)^2 \left[\frac{d r^2}{1 - k r^2} + r^2 d 
\Omega_{d-2}^2 \right] \,,
\end{equation}
which corresponds to the identifications
\begin{equation}
q_{ab}d y^a d y^b = -d t^2 + a(t)^2 \frac{d r^2}{1-k r^2}, \qquad 
\varphi = a(t) r \,,
\end{equation}
with $k=-1,0,+1$ corresponding to open, flat, and closed spatial geometry (for 
the moment we keep arbitrary $d \geq 4$ dimensions).

Within this framework, the full gravitational dynamics can be recast into an 
effective lower-dimensional system, whose structure will allow for a 
remarkable simplification of the cosmological equations, reducing 
the evolution to a relation between a single curvature scalar and the matter 
content of the universe. 

\subsection{Reduced Horndeski action and field equations}

The dimensional reduction described above leads to an effective 
two-dimensional scalar-tensor theory of Horndeski type, involving the metric 
$q_{ab}$ and the scalar field $\varphi$ \cite{Borissova:2026klg}. The most 
general two-dimensional action of this class 
can be written as
\begin{align}
S = \int d^2 y \sqrt{-q} \Big[ 
h_2(\varphi,\chi) 
- h_3(\varphi,\chi) \Box \varphi 
+ h_4(\varphi,\chi)\mathcal{R} \nonumber \\
- 2 \partial_\chi h_4(\varphi,\chi) \left( (\Box \varphi)^2 
- \nabla_a \nabla_b \varphi \nabla^a \nabla^b \varphi \right)
\Big],
\label{action}
\end{align}
where $\mathcal{R}$ is the Ricci scalar of the metric $q_{ab}$ and 
\begin{equation}
\chi = \nabla_a \varphi \nabla^a \varphi
\end{equation}
is the kinetic term of the scalar field. 
This action represents the most general scalar-tensor theory in two dimensions 
leading to second-order equations of motion, ensuring the absence of 
higher-derivative instabilities at the level of the reduced dynamics.

Variation with respect to the metric yields field equations of the 
form \cite{Borissova:2026klg}
\begin{equation}
\mathcal{E}_{ab} = - \frac{1}{2}\left[\alpha + 2\beta \Box \varphi\right] q_{ab}
+ \omega \nabla_a \varphi \nabla_b \varphi
+ \beta \nabla_a \nabla_b \varphi
= T_{ab},
\end{equation}
where the functions $\alpha$, $\beta$, and $\omega$ are given by
\begin{align}
\alpha &= h_2 + \chi \partial_\varphi \left(h_3 - 2 \partial_\varphi h_4\right), 
\\
\beta &= \chi \partial_\chi \left(h_3 - 2 \partial_\varphi h_4\right) 
- \partial_\varphi h_4,
\end{align}
and
\begin{equation}
\omega = \partial_\chi \alpha - \partial_\varphi \beta.
\end{equation}
These functions encode the full structure of the theory and determine how the 
underlying higher-curvature gravitational dynamics manifests at the level of 
the reduced system.

Finally, let us  introduce  matter, assuming a 
perfect-fluid description with energy density $\rho$ and pressure $p$. The 
conservation of the energy-momentum 
tensor leads to the standard continuity equation
\begin{equation}
\dot{\rho} + (d-1)(\rho + p)\frac{\dot{a}}{a} = 0,
\label{conservation}
\end{equation}
which follows from diffeomorphism invariance and remains unchanged with 
respect to the standard cosmological case.

At this stage, the system is still described in terms of the functions
$h_2$, $h_3$, and $h_4$. However, as we show in the following subsection,
the imposition of the integrability condition $\omega=0$ guarantees the
existence of a generating function $\Omega$, providing the first crucial
step toward the subsequent simplification of the cosmological dynamics.

\subsection{Modified Friedmann equations}

As anticipated in the previous subsection, a crucial simplification arises once 
one imposes the consistency of the reduced system in homogeneous and isotropic 
backgrounds. In particular, for FLRW configurations the off-diagonal 
components of the field equations vanish identically in the absence of 
corresponding matter sources, leading to the condition \cite{Borissova:2026klg}
\begin{equation}
\omega = \partial_\chi \alpha - \partial_\varphi \beta = 0.
\label{integrability}
\end{equation}
We stress that condition (\ref{integrability}) is a central structural
requirement of the construction considered here, rather than a secondary
technical assumption. In particular, although the reduced
two-dimensional Horndeski form provides the general framework, the
condition $\omega=0$ does not follow generically from this reduction, nor
from the imposition of FLRW symmetry. Instead, it constitutes an
additional non-trivial integrability condition that selects the
quasi-topological subclass under consideration. While in polynomial
quasi-topological gravities this condition is automatically satisfied,
in the non-polynomial case it represents a genuine restriction on the
allowed reduced functions. In summary, this integrability condition defines 
the quasi-topological subclass considered here and plays a central role in the 
simplification of the dynamics.

 The crucial consequence of the integrability condition
$\omega=0$ is that it guarantees the existence of a generating function
$\Omega(\varphi,\chi)$ such that 
\begin{equation}
\alpha=\partial_\varphi\Omega,
\qquad
\beta=\partial_\chi\Omega .
\end{equation}
 Hence, it is precisely the condition $\omega=0$ that allows the
reduced field equations to be cast into an integrable form. Without this
condition, such a description in terms of a single generating function
would not in general be possible, and the subsequent reduction of the
cosmological dynamics to the one-function form employed below would not
follow. 
 
The important step is to express the system in terms of the scalar quantity 
\cite{Borissova:2026klg}
\begin{equation}
\psi \equiv \frac{1 - \chi}{\varphi^2} = \frac{k + \dot{a}^2}{a^2},
\label{psi1}
\end{equation}
which plays the role of an effective curvature invariant and constitutes the 
natural dynamical variable of the theory.

At this stage, we restrict ourselves to the class of non-polynomial
quasi-topological gravities for which the generating function takes the
form \cite{Borissova:2026wmn,Borissova:2026krh}
\begin{equation}
\Omega(\varphi,\psi) = h(\psi)\, \varphi^{d-1},
\label{hpsiphi}
\end{equation}
where $h(\psi)$ is an arbitrary function. We stress that this ansatz
does not represent the most general non-polynomial quasi-topological
construction, but rather defines the subclass that we investigate in
the present work. Its important advantage is that the gravitational
modifications are encoded in a single function, leading to a particularly
simple and phenomenologically useful cosmological description. More
general non-polynomial curvature quasi-topological gravities may involve
additional functional freedom, for instance through an additional
function $g(\psi)$, while non-polynomial curvature-derivative
quasi-topological gravities can exhibit a more general dependence of the
reduced functions on $\varphi$ and $\chi$ 
\cite{Borissova:2026wmn,Borissova:2026krh}. Such
extensions lie beyond the scope of the present analysis. 

 Hence, using the above relations, the 
equations of motion reduce to a single algebraic 
relation between geometry and matter, namely
\begin{equation}
h(\psi) = \frac{16\pi G}{(d-2)(d-1)} \rho  .
\end{equation}
 
Now, for spatially flat geometry ($k=0$), according to (\ref{psi1}) the 
invariant $\psi$  reduces to
\begin{equation}
\psi = H^2.
\end{equation}
Thus,  the cosmological evolution is governed by 
\begin{equation}
h(H^2) = \frac{16\pi G}{(d-2)(d-1)} \rho,
\label{Fr1}
\end{equation}
which is the modified first Friedmann 
equation. Additionally, the conservation equation for the matter fluid 
(\ref{conservation})
can be integrated for a constant equation-of-state parameter $w$, yielding
\begin{equation}
\rho(a) = \rho_0 \left(\frac{a}{a_0}\right)^{(1-d)(1+w)}.
\end{equation}
Thus, once the function $h(H^2)$ is specified, the cosmological evolution 
follows straightforwardly.
In addition, differentiating the modified Friedmann equation (\ref{Fr1}) with 
respect to 
cosmic time and using the conservation equation (\ref{conservation}), one 
obtains the modified 
Raychaudhuri equation, namely
\begin{equation}
\dot{H} = -   \frac{8\pi G}{(d-2) }     \,\frac{\rho + 
p}{h'(H^2)},
\label{Fr2}
\end{equation}
where a prime denotes differentiation with respect to the argument of the 
function.

In summary, within the class of non-polynomial quasi-topological
gravities considered in this work, one results to the modified
Friedmann equations (\ref{Fr1}) and (\ref{Fr2}), supplemented by the 
conservation
equation (\ref{conservation}). This reveals a remarkable feature of this 
subclass,
namely that the entire background cosmological dynamics is encoded in
a single function $h(H^2)$, which arises from the underlying
gravitational construction. Furthermore, as one can see, the field
equations are of second order, an important advantage compared to
generic higher-curvature theories, where the equations of motion are
typically higher-order differential equations.

Finally, we mention that the Infrared limit of the above constructions is 
  \cite{Borissova:2026klg}
\begin{equation}
h(H^2) \rightarrow H^2,
\end{equation}
in which case  general relativity is recovered.

 \subsection{Inclusion of the cosmological constant}

 We close this section by clarifying the role of the cosmological 
constant
within the present framework. In the cosmological analysis that follows, we
choose to introduce $\Lambda$ through the matter sector, treating it as a
constant vacuum-energy contribution. This prescription is particularly
convenient for our purposes, since it allows us to keep the standard
cosmological-constant contribution explicitly separated from the
non-polynomial quasi-topological modification encoded in $h(\psi)$.

We stress, however, that this separation should not be regarded as a
physically essential distinction. From the viewpoint of the dimensional
reduction of the Einstein-Hilbert action, it is equally natural to
incorporate the cosmological constant into the gravitational sector, in
which case its contribution can be included directly in the function
$h(\psi)$. Thus, the prescription adopted here amounts to a convenient
choice of parametrization and organization of the cosmological equations,
rather than to a restriction imposed by the underlying theory.

At the level of the background cosmological dynamics considered in this
work, the two descriptions are equivalent, provided that the corresponding
constant contribution is consistently identified. Our convention has the
practical advantage of making transparent the distinction between the
standard cosmological-constant term and the effective dark-energy
contribution generated by the non-polynomial quasi-topological
modifications.

Concretely, we consider a total energy density of the form
\begin{equation}
\rho = \rho_m + \rho_r + \rho_\Lambda, 
\end{equation} 
where $\rho_m$, $\rho_r$, and $ \rho_\Lambda = \frac{3\Lambda}{8\pi G},$ 
correspond to matter, radiation, 
and vacuum energy, respectively, where $\Lambda$ is the cosmological constant.

From now on we focus on the standard $d=4$ case. Hence, the modified 
Friedmann equations (\ref{Fr1}),(\ref{Fr2})
become
\begin{eqnarray}
&&h(H^2)=   \frac{8\pi G}{3}   (\rho_m + \rho_r )+ \Lambda\\
&&h'(H^2) \dot{H} = -4\pi G (\rho_m + \rho_r +   p_m + 
p_r ) .
\end{eqnarray}
These can be re-written in the standard form 
 \begin{eqnarray}
&& \!\!\!\!\!\!\!\!\!\!\!\!\!\!\!\!\!\!\!
H^2=\frac{8\pi G}{3} (\rho_m + \rho_r+\rho_{DE})\\
&& \!\!\!\!\!\!\!\!\!\!\!\!\!\!\!\! \!\!\!  H^2+\dot{H} = -\frac{4\pi 
G}{3}(\rho_m+3p_m +
\rho_r+3p_r\nonumber\\
&&
\ \ \ \ \ \ \ \ \ \ \ \ \ \  \, 
+\rho_{DE}+ 3p_{DE}),
\end{eqnarray} 
where we have introduced an effective dark-energy sector with energy density 
and pressure
\begin{eqnarray}
&& \!\!\!\!\rho_{DE} = \frac{3}{8\pi G} \left[ H^2 - h(H^2) + \Lambda \right], 
\label{rhoDE}
\\
&&\!\!\!\! p_{DE} = -\frac{1}{8\pi G} \Big\{ 3\left[H^2 - h(H^2) + 
\Lambda\right]  \nonumber\\
&&
\ \ \ \ \ \ \ \ \ \ \ \ \ \  \, \ \ \   
+ 2\dot{H}\left[1 - h'(H^2)\right] \Big\},
\label{pDE}
\end{eqnarray}
 respectively. 
Additionally, the corresponding dark-energy equation-of-state parameter is 
given by    
 \begin{eqnarray}
  w_{DE}\equiv \frac{p_{DE}}{\rho_{DE}}.
 \end{eqnarray}
 Hence, as we observe, within the framework of non-polynomial 
quasi-topological gravity, we obtain an effective dark energy of geometrical 
origin, quantified by a single function. We finally mention that the 
inclusion of an explicit cosmological constant
is not obligatory,   and that its separation from the geometrical
contribution in the above expressions reflects the convention adopted in
the present analysis.

 \section{Viable models and their constructions}

 Having established that,  for the subclass  of non-polynomial
quasi-topological gravities considered here, the background cosmological
dynamics is encoded in a single function $h(H^2)$, we now proceed to
construct explicit realizations and examine the conditions under which
they lead to a viable cosmological evolution.

The algebraic structure of the modified Friedmann equations provides a 
remarkably efficient framework, namely once the function $h(H^2)$ is specified, 
the 
full expansion history follows directly, without the need to solve higher-order 
differential equations. This allows for a systematic exploration of the space 
of models, as well as a direct connection between the underlying gravitational 
theory and observable cosmology.

In this section, we first determine the general requirements that the
function $h(H^2)$ must satisfy in order to recover the standard
radiation- and matter-dominated epochs and yield a consistent late-time
behavior.  We then construct 
explicit models arising within the non-polynomial quasi-topological framework, 
which  provide controlled deviations from the 
$\Lambda$CDM paradigm, while preserving the theoretical consistency and 
simplicity of the underlying construction.

 \subsection{General requirements for viability}

  For the class  of non-polynomial quasi-topological gravity models
considered here, the entire background evolution is determined by the
function $h(H^2)$.  Therefore, 
cosmological viability can be directly translated into conditions on its 
functional form.  A realistic cosmological scenario must recover the
standard radiation-
and matter-dominated eras and subsequently lead to a phase of
accelerated expansion.  In addition, the
resulting dynamics must remain free from pathological behaviors, ensuring a 
consistent and well-defined evolution.

It is convenient to parametrize deviations from general relativity through
\begin{equation}
h(H^2) = H^2 + f(H^2),
\end{equation}
where $f(H^2)$ encodes genuine modifications of the gravitational sector.  As 
we mentioned in the previous section, general relativity is recovered in the 
limit $f(H^2)=0$, while non-trivial 
forms of $f(H^2)$ describe departures from $\Lambda$CDM cosmology.  

In order to recover the standard radiation and matter eras, the modified 
Friedmann equation must reduce effectively to its standard form, implying
\begin{equation}
H^2 \simeq \frac{8\pi G}{3}\rho,
\end{equation}
during these epochs. This requires that the modification term remains 
subdominant in the corresponding regimes, namely
\begin{equation}
\frac{f(H^2)}{H^2} \ll 1,
\end{equation}
for the range of $H$ relevant to radiation and matter domination. Hence,
the ratio $f(H^2)/H^2$ must remain sufficiently small over
the range of $H$ relevant to the standard radiation- and
matter-dominated epochs.
Furthermore, at late times, accelerated expansion can be driven by the combined 
effect of cosmological 
constant  and  the modification term itself. The 
condition for acceleration can be expressed through the deceleration 
parameter
\begin{equation}
q = -1 - \frac{\dot{H}}{H^2},
\label{deceleration}
\end{equation}
which must become negative. Using the modified Friedmann equations, this 
translates into constraints on the form of $h(H^2)$ and its derivative.
Finally, general consistency requirements must also be satisfied, namely 
$h(H^2)$ must be non-negative and the solutions must correspond 
to real and positive values of $H^2$, ensuring a physically meaningful 
cosmological evolution.

Summarizing, cosmological viability in non-polynomial quasi-topological 
gravity requires that the function $h(H^2)$: (i) reproduces the standard 
radiation and matter eras, (ii) leads to late-time acceleration, and (iii) 
satisfies basic consistency conditions such as monotonicity and positivity.

 In the following, we construct explicit realizations within this
framework and investigate their cosmological properties, focusing in
particular on their late-time phenomenology.

\subsection{Polynomial model}

We now demonstrate explicitly that modified Friedmann equations of the form  
(\ref{Fr1}),(\ref{Fr2}), with
\begin{equation}
h(H^2) = \sum_{n=1}^{N} b_n H^{2n},  
\label{model11}
\end{equation}
can be consistently obtained from an appropriate choice of the underlying 
reduced   functions $h_2(\varphi,\chi)$, $h_3(\varphi,\chi)$, and 
$h_4(\varphi,\chi)$ of the action (\ref{action}). This establishes a direct 
link between the fundamental 
gravitational theory and phenomenological cosmological models.

Following the general procedure of Section \ref{NPQTGtheory}, we introduce the 
scalar variable
\begin{equation}
\psi = \frac{1-\chi}{\varphi^2},
\end{equation}
and define the generating function as
\begin{equation}
\Omega(\varphi,\psi) = \varphi^{3} h(\psi).
\end{equation}
For a polynomial form of $h(\psi)$, this yields
\begin{equation}
\Omega(\varphi,\chi) 
= \sum_{n=1}^{N} b_n \, \varphi^{3-2n}(1-\chi)^n,
\end{equation}
which by construction satisfies the integrability condition 
$\partial_\chi \alpha - \partial_\varphi \beta = 0$. Thus, the corresponding 
functions $\alpha$ and $\beta$ are then obtained as
\begin{equation}
\alpha(\varphi,\chi) = \partial_\varphi \Omega
= \sum_{n=1}^{N} (3-2n)b_n \, \varphi^{2-2n}(1-\chi)^n,
\end{equation}
\begin{equation}
\beta(\varphi,\chi) = \partial_\chi \Omega
= -\sum_{n=1}^{N} nb_n \, \varphi^{3-2n}(1-\chi)^{n-1}.
\end{equation}

 To provide an explicit realization of the reduced action, we consider for
simplicity the convenient choice 
\begin{equation}
h_4(\varphi,\chi)=0,
\end{equation}
in which the defining relations reduce to
\begin{equation}
\alpha = h_2 + \chi\,\partial_\varphi h_3,
\qquad
\beta = \chi\,\partial_\chi h_3.
\end{equation}
The function $h_3$ can then be determined from
\begin{equation}
\partial_\chi h_3 = \frac{\beta}{\chi}
= -\sum_{n=1}^{N} nb_n \, \varphi^{3-2n}
\frac{(1-\chi)^{n-1}}{\chi},
\end{equation}
which upon integration yields
\begin{equation}
h_3(\varphi,\chi)
=
-\sum_{n=1}^{N} nb_n \, \varphi^{3-2n}
\int^\chi \frac{(1-u)^{n-1}}{u}\,du,
\end{equation}
up to an arbitrary function of $\varphi$. For integer $n$, the integral can be 
evaluated explicitly, leading to logarithmic and polynomial contributions in 
$\chi$. The function $h_2$ is then obtained through
\begin{equation}
h_2(\varphi,\chi)
=
\alpha(\varphi,\chi) - \chi\,\partial_\varphi h_3(\varphi,\chi).
\end{equation}

By construction, the generating function takes the form 
$\Omega=\varphi^{3}h(\psi)$, and therefore the cosmological reduction 
directly leads to (\ref{model11}). Hence, by introducing this form to the 
effective dark energy density and pressure (\ref{rhoDE}),(\ref{pDE})
leads to 
 \begin{eqnarray}
&& \! \! \! \! \!\!\!\! \! \! \! \! \!\rho_{DE} = \frac{3}{8\pi G} \left[ H^2 - 
\sum_{n=1}^{N} b_n H^{2n} + \Lambda \right], 
\label{rhoDE11}
\\
&&\!\!\!\!  \! \! \! \! \! \! \! \! \!p_{DE} = -\frac{1}{8\pi G} \left[ 
3\left(H^2 -\sum_{n=1}^{N} b_n H^{2n} + 
\Lambda\right) \right.\nonumber\\
&&
\ \ \ \ \ \ \ \ \ \ \ \ \      \left.
+ 2\dot{H}\left(1 -\sum_{n=1}^{N} n  b_n H^{2(n-1)}\right) \right].
\label{pDE11}
\end{eqnarray}

This construction demonstrates that polynomial modifications of the 
Friedmann equation can be systematically embedded within the 
non-polynomial quasi-topological framework (note that at the level of 
energy density, these models are similar to those obtained within the running 
vacuum constructions  
\cite{Basilakos:2012ra,Sola:2016jky,SolaPeracaula:2021gxi}). Different choices 
of the 
coefficients $b_n$ correspond to different higher-curvature corrections, 
while the integrability condition is automatically ensured through the 
existence of the generating function $\Omega$. 
 Finally, we stress that the reconstruction at the level of the reduced 
functions $h_2$, $h_3$, and $h_4$ is not unique. Indeed, the quantities 
that determine the reduced field equations are the combinations $\alpha$ 
and $\beta$, or equivalently, once the integrability condition is imposed, 
the generating function $\Omega$. Consequently, different choices of 
$(h_2,h_3,h_4)$ that reproduce the same $\alpha$ and $\beta$, and hence 
the same generating function $\Omega$, lead to the same background 
equations and cosmological solutions. The explicit functions constructed 
above should therefore be regarded as one convenient realization of the 
desired function $h(H^2)$, rather than as a unique action-level 
reconstruction. Alternative realizations of the reduced functions $h_i$ 
have been discussed in Refs. \cite{Borissova:2026dlz,Borissova:2026rbi}. Thus, 
for the background analysis 
performed here, the particular realization of the $h_i$ is not essential, 
provided that the corresponding physical combinations $\alpha$, $\beta$, 
or equivalently $\Omega$, remain unchanged.

\subsection{Quartic model}
\label{Quartic}

It proves interesting to focus on the quartic sub-class of the above models. 
 Such a construction illustrates how simple and 
phenomenologically relevant modifications of the Friedmann equation arise 
naturally within non-polynomial quasi-topological gravity.

According to the above analysis, we consider the following choice of the 
reduced Horndeski functions:
\begin{equation}
h_4(\varphi,\chi)=0,
\end{equation}
\begin{equation}
h_3(\varphi,\chi)= -\left(\varphi + \frac{2b}{\varphi}\right)\ln|\chi| 
+ \frac{2b}{\varphi}\chi,
\end{equation}
\begin{equation}
h_2(\varphi,\chi)=1-\chi+\chi\ln|\chi| 
+ \frac{b}{\varphi^2}\left(-1+2\chi+\chi^2-2\chi\ln|\chi|\right),
\end{equation}
where $b$ is a constant parameter controlling the deviation from general 
relativity.

Since $h_4=0$, the functions $\alpha$ and $\beta$ simplify to
$
\alpha = h_2 + \chi \partial_\varphi h_3$ and $
\beta = \chi \partial_\chi h_3,
$
and a direct calculation yields
\begin{equation}
\alpha = (1-\chi) - \frac{b}{\varphi^2}(1-\chi)^2,
\end{equation}
\begin{equation}
\beta = -\varphi - \frac{2b}{\varphi}(1-\chi).
\end{equation}
These expressions satisfy the integrability condition, ensuring the existence 
of a generating function $\Omega(\varphi,\chi)$ such that
$
\alpha = \partial_\varphi \Omega,
$ and $
\beta = \partial_\chi \Omega$,
and integrating  we obtain
\begin{equation}
\Omega(\varphi,\chi) = \varphi(1-\chi) + \frac{b}{\varphi}(1-\chi)^2.
\end{equation}
Now, introducing    
\begin{equation}
\psi = \frac{1-\chi}{\varphi^2},
\end{equation}
the generating function can be written as
\begin{equation}
\Omega = \varphi^3\left(\psi + b\psi^2\right),
\end{equation}
from which we immediately identify
\begin{equation}
h(\psi) = \psi + b\psi^2.
\end{equation}
Since for spatially flat cosmology  $\psi=H^2$,  we straightforwardly obtain
\begin{equation}
h(H^2 ) = H^2  + bH^4.
\label{quartic1}
\end{equation}
Finally,  introducing this form to the 
effective dark energy density and pressure (\ref{rhoDE}),(\ref{pDE})
we acquire
 \begin{eqnarray}
&& \! \! \! \! \!\!\!\! \! \! \! \! \!\rho_{DE} = \frac{3}{8\pi G} \left(   
 \Lambda-bH^4  \right), 
\label{rhoDE11}
\\
&&\!\!\!\!  \! \! \! \! \! \! \! \! \!p_{DE} = -\frac{1}{8\pi G} \left[ 
3\left(
\Lambda-bH^4 \right)  
-4b\dot{H}  H^2\right],
\label{pDE11}
\end{eqnarray} 
and thus    the corresponding dark-energy equation-of-state parameter 
will be  
 \begin{eqnarray}
  w_{DE}\equiv -1 + \frac{     
   4b\dot{H}  H^2 }{ {3}  \left(   
 \Lambda-bH^4  \right)}.
 \label{wdequartic}
 \end{eqnarray}

This model therefore provides a minimal extension of $\Lambda$CDM paradigm, in 
which higher-curvature effects manifest as a quartic correction in $H^2$, while 
the dynamical structure of the equations remains second-order and algebraic. In 
the limit $b \to 0$, standard cosmology is recovered, whereas for non-zero 
$b$ the model introduces controlled deviations that can be constrained by 
observations.

We emphasize that the simple quartic realization considered here is
employed as an effective description of the late-time cosmological
evolution. Indeed, the relative contribution of the quartic correction
with respect to the Einstein term scales as $bH^2$, and therefore does
not generically become negligible towards increasingly high curvature.
Consequently, we do not interpret this simple quartic realization as a
complete description of the radiation-dominated and still earlier
high-curvature regimes, which may require additional terms in
$h(H^2)$.
 
 \subsection{Power-law model}
 \label{Powermode}

Another interesting class of models is the  power-law modification of the form
\begin{equation}
h(H^2) = H^2 + b H^{\delta},
\label{powerlaw}
\end{equation}
where $b$ is a constant parameter and $\delta$ controls the deviation from 
general relativity.  

In terms of the invariant $\psi$, this corresponds to
\begin{equation}
h(\psi) = \psi + b \psi^{\delta/2}.
\end{equation}
Following the general construction of Section  \ref{NPQTGtheory}, we start 
from 
\begin{equation}
\Omega(\varphi,\psi) = \varphi^{3} h(\psi),
\end{equation}
which in terms of $(\varphi,\chi)$ becomes
\begin{equation}
\Omega(\varphi,\chi) 
= \varphi(1-\chi) 
+ b\, \varphi^{3-\delta}(1-\chi)^{\delta/2},
\end{equation}
which satisfies the integrability condition and ensures 
the 
consistency of the reduced system. The corresponding functions $\alpha$ and 
$\beta$ 
are obtained from
$
\alpha = \partial_\varphi \Omega$ and $\beta = \partial_\chi \Omega$
leading to
\begin{equation}
\alpha = (1-\chi) 
+ b(3-\delta)\varphi^{2-\delta}(1-\chi)^{\delta/2},
\end{equation}
\begin{equation}
\!\!\!\!\!\!\!\!\!\!\!\!\!\!\!\!\!\!\!\!\!\!\!\!\!\!\!\!\!\!\!\!\!\!\!\!
\beta = -\varphi 
- b\,\frac{\delta}{2}\,\varphi^{3-\delta}(1-\chi)^{\frac{\delta}{2}-1}.
\end{equation}

In order to  provide an explicit realization of the reduced action, we 
again
adopt for simplicity the convenient choice $h_4(\varphi,\chi)=0$, such that
$\alpha=h_2+\chi\partial_\varphi h_3$ and
$\beta=\chi\partial_\chi h_3$.  Hence, the function $h_3$ is then determined 
from
\begin{equation}
\partial_\chi h_3 = \frac{\beta}{\chi}
= -\frac{\varphi}{\chi}
- b\,\frac{\delta}{2}\,\varphi^{3-\delta}
\frac{(1-\chi)^{\frac{\delta}{2}-1}}{\chi},
\end{equation}
which upon integration yields
\begin{equation}
h_3(\varphi,\chi)
=
-\varphi \ln|\chi|
- b\,\frac{\delta}{2}\,\varphi^{3-\delta}
\int^\chi \frac{(1-u)^{\frac{\delta}{2}-1}}{u}\,du,
\end{equation}
up to an arbitrary function of $\varphi$.
Finally, the function $h_2$ is obtained through
\begin{equation}
h_2(\varphi,\chi)
=
\alpha(\varphi,\chi) - \chi\,\partial_\varphi h_3(\varphi,\chi).
\end{equation}

In summary, with the above function choices, we obtain  the power-law 
deformation (\ref{powerlaw}). Hence, by inserting this form to the 
effective dark energy density and pressure (\ref{rhoDE}),(\ref{pDE})
leads to  
\begin{eqnarray}
&& \!\!\!\! \!\!\!\!\!\!\!\!\!\!\!\!\!\!\!\rho_{DE} = \frac{3}{8\pi G} \left( 
\Lambda -  b H^{\delta}   \right), 
\label{rhoDE11b}
\\
&&\!\!\!\! \!\!\!\!\!\!\!\!\!\!\!\!\!\!\! p_{DE} = -\frac{1}{8\pi G} \left\{ 
3\left(\Lambda -  b H^{\delta} \right) 
- b \delta \dot{H}   H^{\delta-2} \right\},
\label{pDE11b}
\end{eqnarray}
and therefore     the   dark-energy equation-of-state parameter 
reads as  
 \begin{eqnarray}
  w_{DE}\equiv -1 + \frac{  b \delta \dot{H}   H^{\delta-2} }{ {3}  
\left(   
 \Lambda- b H^{\delta}  \right)}.
 \label{wdepower}
 \end{eqnarray}
 
The Power-law model provides a natural extension of the quartic 
scenario, allowing for a continuous deformation of the higher-curvature 
correction  (note that at the level of 
energy density, these models are similar to those obtained within 
power-law $f(T)$ gravity \cite{Cai:2015emx} or entropic constructions
\cite{Luciano:2025hjn,Luciano:2026ufu,Leizerovich:2026pfy,Saridakis:2026vxo}).
In the limit $\delta \to 4$ the Quartic model is recovered, while for 
general $\delta$ the modification can become either milder or stronger, 
depending on the energy scale. Lastly, for $\delta=0$, $\Lambda$CDM 
scenario is recovered, even in the case where we do not consider an explicit 
cosmological constant $\Lambda$.

\subsection{Non-polynomial models}

The formalism of non-polynomial quasi-topological gravity naturally allows 
for a wide class of non-polynomial modifications of the Friedmann equation, 
extending beyond the simple polynomial corrections presented above. In this 
subsection we provide a representative example, illustrating how such models 
can 
be consistently embedded within the underlying gravitational framework.

We are interested in obtaining the function
\begin{equation}
h(H^2) = \frac{H^2}{1 - \left(\frac{H^2}{H_*^2}\right)^\gamma},
\end{equation}
where $H_*$ is a characteristic energy scale and $\gamma$ is a positive 
parameter controlling the deviation from general relativity. This form 
introduces a non-polynomial dependence on $H^2$, which becomes significant 
at high curvature, while reducing to the standard behavior at low energies.

In terms of the invariant $\psi$, this corresponds to
\begin{equation}
h(\psi) = \frac{\psi}{1 - \left(\frac{\psi}{H_*^2}\right)^\gamma}.
\end{equation}
Hence, following the general construction presented in Section 
\ref{NPQTGtheory}, we define the generating function
\begin{equation}
\Omega(\varphi,\psi) = \varphi^3 h(\psi),
\end{equation}
which in terms of $(\varphi,\chi)$ reads
\begin{equation}
\Omega(\varphi,\chi) =
\frac{\varphi(1-\chi)}
{1 - \left[\dfrac{1-\chi}{\varphi^2 H_*^2}\right]^\gamma}.
\end{equation}
By construction, this form satisfies the integrability condition and thus 
ensures the consistency of the reduced theory.

An explicit realization of the corresponding reduced action (\ref{action}) can 
be obtained 
by choosing
\begin{equation}
h_4(\varphi,\chi)=0,
\end{equation}
and reconstructing the remaining functions from
$
\beta = \partial_\chi \Omega
$ and $\alpha = \partial_\varphi \Omega$,
together with
$\beta = \chi\,\partial_\chi h_3$ and 
$\alpha = h_2 + \chi\,\partial_\varphi h_3$.
Now, introducing
\begin{equation}
u = \left(\frac{1-\chi}{\varphi^2 H_*^2}\right)^\gamma,
\end{equation}
we obtain
\begin{equation}
\partial_\chi h_3(\varphi,\chi)
=
-\frac{\varphi}{\chi}\,
\frac{1 + (\gamma-1)u}{(1-u)^2},
\end{equation}
and therefore the function $h_3$ can be written explicitly as
\begin{equation}
h_3(\varphi,\chi)
=
-\varphi \int^\chi
\frac{1 + (\gamma-1)\left(\dfrac{1-u'}{\varphi^2 H_*^2}\right)^\gamma}
{u'\left[1 - \left(\dfrac{1-u'}{\varphi^2 H_*^2}\right)^\gamma\right]^2}
\,du',
\end{equation}
up to an arbitrary function of $\varphi$. Finally, the function $h_2$ then 
follows from
\begin{equation}
h_2(\varphi,\chi)
=
\alpha(\varphi,\chi) - \chi\,\partial_\varphi h_3(\varphi,\chi).
\end{equation}

Although these expressions are more involved than in the polynomial case, 
they provide an explicit and consistent embedding of the above 
non-polynomial function $h(H^2)$ within the reduced Horndeski framework.
In particular,  inserting  this form to the 
effective dark energy density and pressure (\ref{rhoDE}),(\ref{pDE})
we obtain
\begin{eqnarray}
&& \!\!\!\!\!\!\!\!\!\!\!\!\!\!\!\rho_{DE} = \frac{3}{8\pi G} \left[ H^2 
-\frac{H^2}{1 - \left(H^2/H_*^2\right)^\gamma} + \Lambda \right], 
\label{rhoDE22}
\\
&& \!\!\!\!\!\!\!\!\! \!\!\!\!\!\!p_{DE} = -\frac{1}{8\pi G} \left\{ 3\left[H^2 
-\frac{H^2}{1 - \left(H^2/H_*^2\right)^\gamma} + 
\Lambda\right] \right.\nonumber\\
&&
\ \ \ \ \,          \left.
+ 2\dot{H}\left[1 - \frac{
1 + (\gamma - 1)\left(H^2/H_*^2\right)^\gamma
}{
\left[1 - \left(H^2/H_*^2\right)^\gamma\right]^2
}\right] \right\},
\label{pDE22}
\end{eqnarray}
from which we can obtain the corresponding equation-of-state parameter.
At low energies, $H^2 \ll H_*^2$, one recovers $h(H^2)\simeq H^2$, ensuring the 
standard cosmological behavior. However, at high curvature the 
non-polynomial structure introduces strong deviations, which  modify the 
early-time dynamics. 

We close this section by mentioning that  all the above constructions preserve 
the algebraic structure of the  theory, which constitutes one of 
the main advantages of    non-polynomial quasi-topological gravity.

\section{Cosmological evolution and observational constraints}

 In the previous section we illustrated the flexibility of the non-polynomial 
quasi-topological framework. Different forms of $h(H^2)$ correspond to 
different physical scenarios, ranging from small deviations from $\Lambda$CDM 
to strongly modified high-curvature dynamics, and  
 once the function $h(H^2)$ is specified, the full cosmological 
evolution can be determined analytically, allowing the computation of 
observables such as  the Hubble parameter  and the dark-energy 
equation-of-state 
parameter.

In this section we will proceed to the investigation of specific cosmological 
scenarios in non-polynomial quasi-topological gravity. As   representative 
examples we choose the quartic model and the power-law model presented in 
subsections \ref{Quartic} and \ref{Powermode} respectively,  and focus on
their late-time cosmological evolution and observational constraints.

  In order to confront the proposed models with observational data, we consider 
a combination of complementary late-time cosmological probes. In particular, 
we employ Type Ia Supernovae (SNIa), Cosmic Chronometer (CC) measurements of 
the Hubble parameter $H(z)$, and Baryon Acoustic Oscillation (BAO) data.

Our analysis is performed within a  Bayesian inference  framework, 
adopting a total likelihood of the form 
$\mathcal{L}_{\textrm{tot}} \propto \exp(-\chi^2_{\textrm{tot}}/2)$, where 
the total chi-square function is constructed as a sum of the individual 
contributions,
\begin{equation}
\chi^2_{\textrm{tot}} = \chi^2_{\textrm{SNIa}} + \chi^2_{\textrm{BAOs}} + 
\chi^2_{\textrm{CC}}.
\end{equation}
 The SNIa and CC likelihoods follow the standard methodology presented in
\cite{Anagnostopoulos:2019miu}.  More specifically, for the supernova sector 
we employ the 40-point binned Pantheon sample. The corresponding contribution 
to the likelihood is evaluated using the full covariance matrix of the binned 
sample, according to
\begin{equation}
\chi^2_{\rm SNIa}
=
\Delta {\bf m}^{\,T}
{\bf C}_{\rm SNIa}^{-1}
\Delta {\bf m},
\label{chiSNIa}
\end{equation}
where $\Delta {\bf m}={\bf m}_{\rm obs}-{\bf m}_{\rm th}$ denotes the 
difference between the observed and theoretical apparent magnitudes, while 
${\bf C}_{\rm SNIa}$ is the corresponding total covariance matrix. In 
particular, the provided covariance matrix is supplemented by the individual 
statistical magnitude uncertainties through the corresponding diagonal 
contributions. 

For the BAO sector, we utilize the dataset of \cite{BOSS:2016wmc}, which 
incorporates fiducial cosmology corrections through the ratio 
$r_{\textrm{d}}/r_{\textrm{fid}}$, allowing for a consistent comparison 
between theoretical predictions and observations.  Thus, the combined 
dataset considered in the present analysis consists of 
40 SNIa, 31 CC, and 6 BAO measurements, corresponding to a total of 
$N=77$ data points. 

The exploration of the parameter space is carried out using the 
 Markov Chain Monte Carlo (MCMC)  method, implemented through the 
publicly available Python package \texttt{emcee} \cite{ForemanMackey:2012ig}. 
In this context, the SNIa nuisance parameter $\mathcal{M}$ is treated as a free 
parameter of the Pantheon dataset (see \cite{Anagnostopoulos:2019miu} and 
references therein), while the radiation density parameter $\Omega_{r0}$ is 
neglected, as it is subdominant at late times.

The above statistical framework is applied to the quartic and power-law models, 
as well as to the standard $\Lambda$CDM scenario, allowing 
for a direct and consistent comparison between the two. In all cases, we 
employ $1000$ walkers and $2500$ steps per walker, assuming flat priors for 
all parameters.

 \subsection{Quartic model }

 Let us focus on the Quartic model (\ref{quartic1}), which gives rise to the 
effective dark energy density and pressure  (\ref{rhoDE11}),(\ref{pDE11}), 
respectively.  We focus on 
the dust-matter case, i.e. 
we set $p_m=0$.
Furthermore, we introduce the 
density parameters 
 \begin{eqnarray} \label{FRWomatter}
&&\Omega_m=\frac{8\pi G}{3H^2} \rho_m\\
&& \label{FRWode}
\Omega_{DE}=\frac{8\pi G}{3H^2} \rho_{DE},
 \end{eqnarray} 
 and we use the subscript ``0" to mark  the present   value of a quantity.
 Additionally,  as the independent variable  we use the redshift  $ 
1+z=a_0/a$ and we set the current scale factor $a_0=1$. 
In this case,  
the modified Friedmann equation (\ref{Fr1}) can be solved algebraically 
to give
\begin{equation}
H^2(z) = \frac{-1 + \sqrt{1 + 4b \left[\frac{8\pi G}{3}\rho_m(z) + \Lambda 
\right]}}{2b},
\end{equation}
where we have selected the branch that recovers general relativity in the limit 
$b \to 0$.
For negative $b$, the reality of this branch additionally requires
\begin{equation}
1+4b\left[\frac{8\pi G}{3}\rho_m(z)+\Lambda\right]\geq 0 ,
\end{equation}
and therefore the solution cannot in general be extended to arbitrarily
high redshift. This further motivates treating the quartic realization
as an effective late-time cosmological model in the present analysis.

Moreover, 
we will use the standard evolution
$
\rho_m(z)=\rho_{m0}(1+z)^3$ that arises from (\ref{conservation}) in the   
case of dust matter. Hence, knowing $H^2(z) $ and $
\rho_m(z)$ we can calculate the density parameters $\Omega_m(z)$, 
$\Omega_{DE}(z)$, as well as the dark-energy equation-of-state parameter  
 $w_{DE}(z)$ from (\ref{wdequartic}) and the deceleration parameter from 
 (\ref{deceleration}).

    \begin{figure}[!]
        \vspace{-0.8cm}
           \hspace{-0.6cm}
  \includegraphics[scale=0.4]{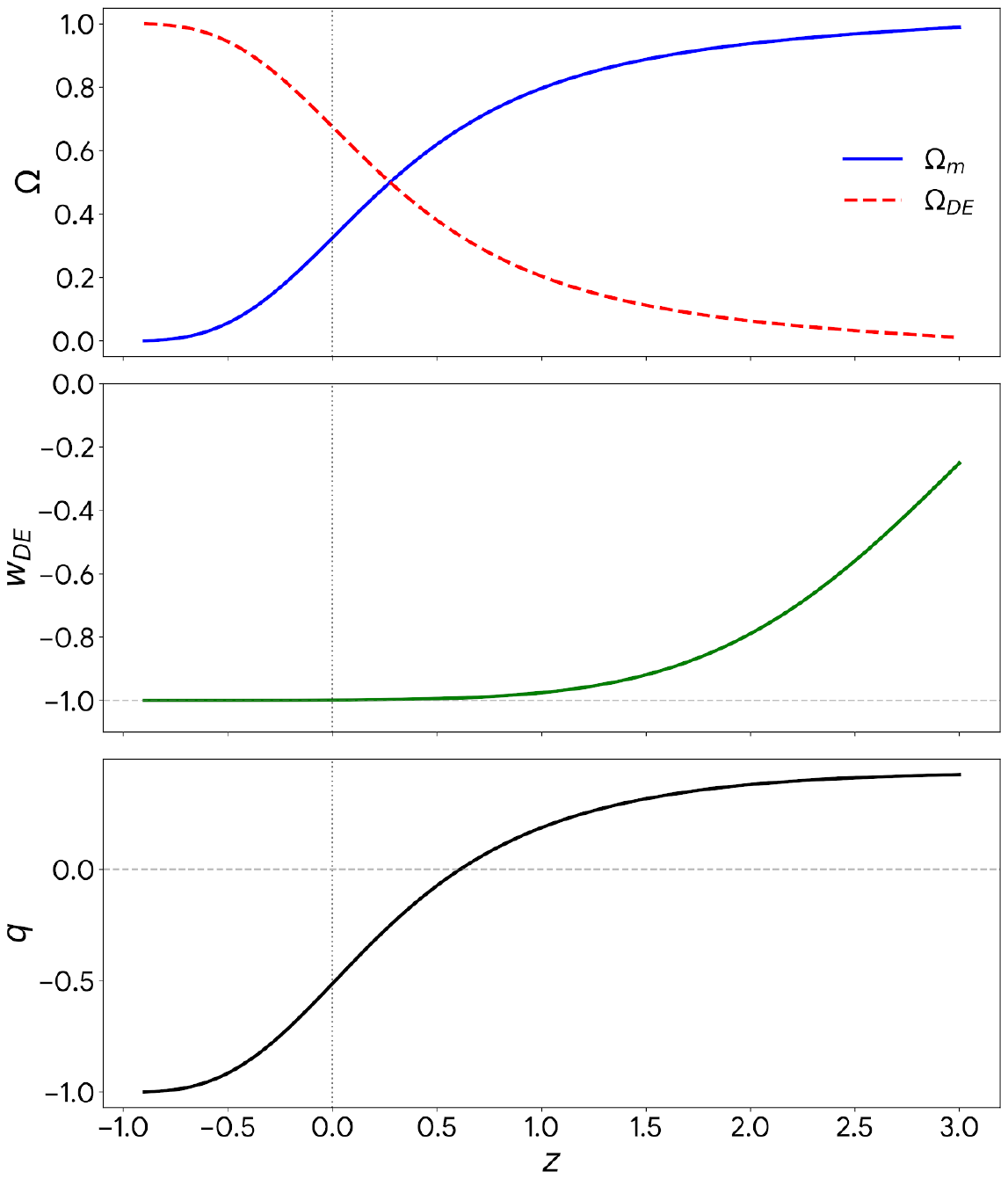}
    \vspace{-1.7cm}
\caption{
{\it{ Upper panel: The evolution of   the matter density
parameter $\Omega_{m}$ (blue-solid) and of the 
dark energy    
density parameter $\Omega_{DE}$  (red-dashed), 
of the Quartic model (\ref{quartic1}), 
as a function of the redshift $z$,  for
 $b=-0.01$  in units 
where $H_0=1$.
 Middle panel: The evolution of the corresponding dark
energy equation-of-state parameter $w_{DE}$ from  (\ref{wdequartic}). Lower 
panel:  The evolution of the 
corresponding   deceleration parameter $q$ from (\ref{deceleration}). In all 
graphs 
we   impose 
$\Omega_{DE}(z=0)\equiv\Omega_{DE0}\approx0.69$  at present, and we 
have added a vertical dotted line denoting 
the current time $z=0$.
}} }
\label{Omegasquartic}
\end{figure}

 \begin{figure}[ht]
 \hspace{-0.7cm}
\includegraphics[scale=0.3]{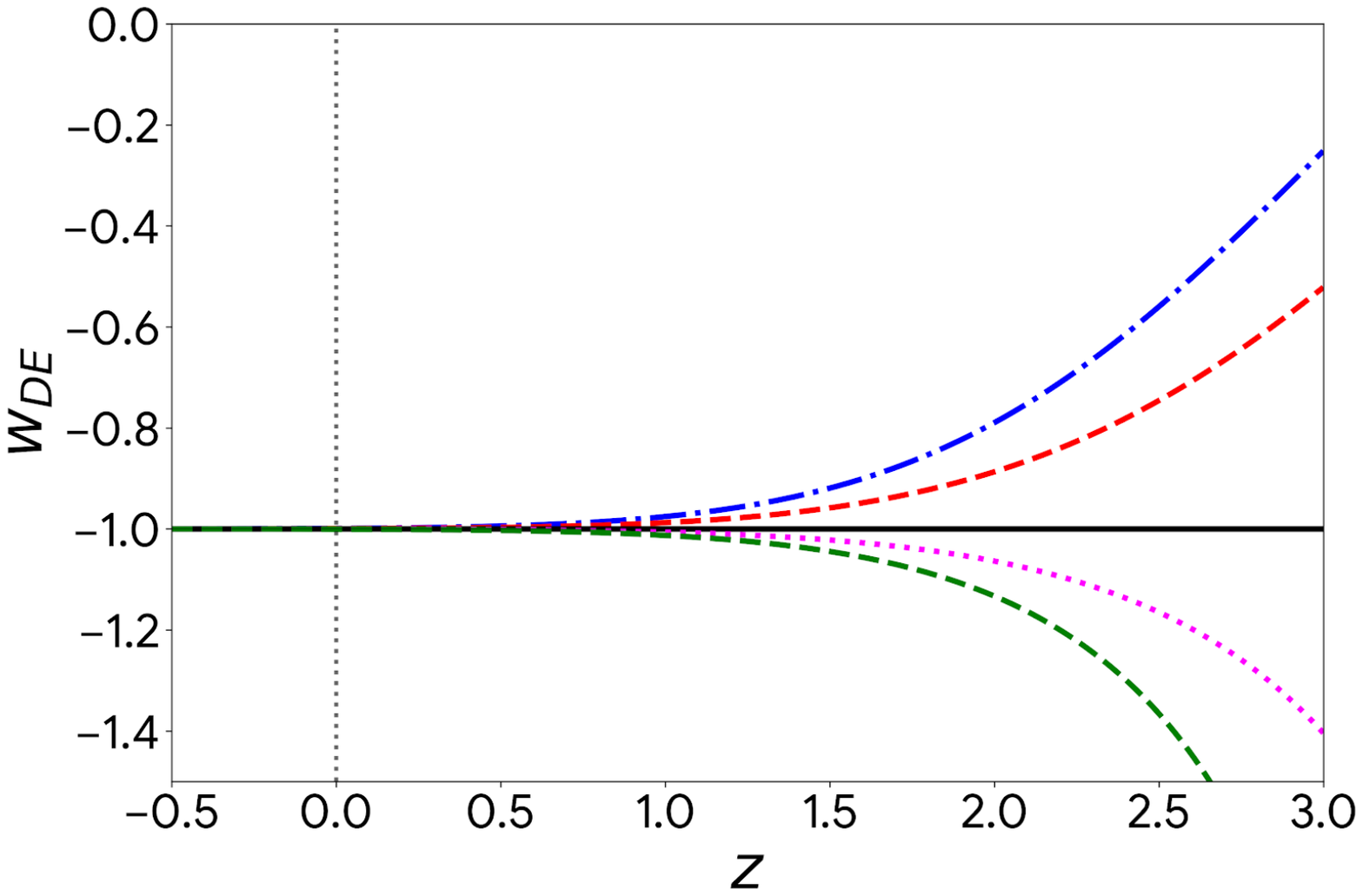}
\caption{
{\it{The evolution of the equation-of-state parameter $w_{DE}$ of   the Quartic 
model (\ref{quartic1}), 
as a function of the redshift $z$,    for 
$b=0$  (black-solid), $b=-0.01$ (blue-dashed-dotted),  $b=-0.005$
(red-dashed),  $b=+0.0025$ (pink-dotted)
 and  $b=+0.005$ 
(green-dashed), in units 
where $H_0=1$.
In all graphs we have imposed 
$\Omega_{DE}(z=0)\equiv\Omega_{DE0}\approx0.69$ at present,  and we 
have added a vertical dotted line denoting 
the current time $z=0$.}} }
\label{wmultiplotquartic}
\end{figure}

 \begin{table*}[t]
\centering
\begin{tabular}{lccccccc}
\hline\hline
Model & $\Omega_{m0}$ & $h$ & $r_d$ & $B=b H_0^2$ & $\mathcal{M}$ & 
$\chi^2_{\min}$ & $\chi^2_{\min}/\mathrm{dof}$ \\
\hline
Quartic NPQTG 
& $0.3243^{+0.0413}_{-0.0312}$ 
& $0.6871^{+0.0181}_{-0.0177}$ 
& $146.557^{+3.706}_{-3.596}$ 
& $0.0153^{+0.0238}_{-0.0406}$ 
& $-19.3849^{+0.0534}_{-0.0540}$ 
& $57.097$ 
& $0.793$ \\
\hline\hline
\end{tabular}
\caption{Observational constraints for the Quartic   model (\ref{quartic1}) 
from 
the joint 
SNIa/CC/BAO analysis, together with the corresponding minimum chi-square and 
reduced chi-square. Here $\mathrm{dof}$ denotes the number of degrees of 
freedom, defined as the number of data points minus the number of fitted 
parameters.}
\label{tab:quartic}
\end{table*}
 
In the upper panel of Fig.~\ref{Omegasquartic} we present the evolution of the 
density parameters,  as 
functions of redshift. Moreover, the middle panel displays the corresponding 
behavior of   $w_{DE}(z)$, 
while the lower panel illustrates the evolution of the deceleration parameter.
We have fixed the present-day density 
parameters to $\Omega_{DE}(z=0)\equiv\Omega_{DE0}\simeq 0.69$ and 
$\Omega_m(z=0)\equiv\Omega_{m0}\simeq 0.31$, in agreement with observational 
constraints \cite{Planck:2018vyg}, and for completeness, the evolution is 
extended into the future regime, approaching the limit $z \rightarrow -1$.
As can be seen, within the late-time regime considered here, the
scenario exhibits the expected sequence of matter domination followed
by dark-energy domination. At late times, the evolution asymptotically
approaches a fully dark-energy-dominated phase.   Furthermore,
the transition from decelerated to accelerated expansion occurs at 
$z\approx 0.6$, in agreement with current observational constraints. 

Concerning the effective dark-energy equation-of-state parameter, we observe 
that its present value is close to $w_{DE}(z=0)\simeq -1$, consistently with 
observations, while at the same time exhibiting a non-trivial dynamical 
evolution, as discussed above.

Motivated by this dynamical behavior, it is instructive to examine the 
dependence of $w_{DE}$ on the model parameters. In Fig.~\ref{wmultiplotquartic} 
we present 
the evolution of $w_{DE}(z)$ for various values of $b$. As 
expected, for $b=0$ the model reduces to the standard $\Lambda$CDM 
scenario. Negative $b$ values drives  $w_{DE}$ 
into the 
quintessence regime, whereas positive $b$ brings the model into the phantom 
regime.

\begin{figure}[!]
\includegraphics[width=0.48\textwidth]{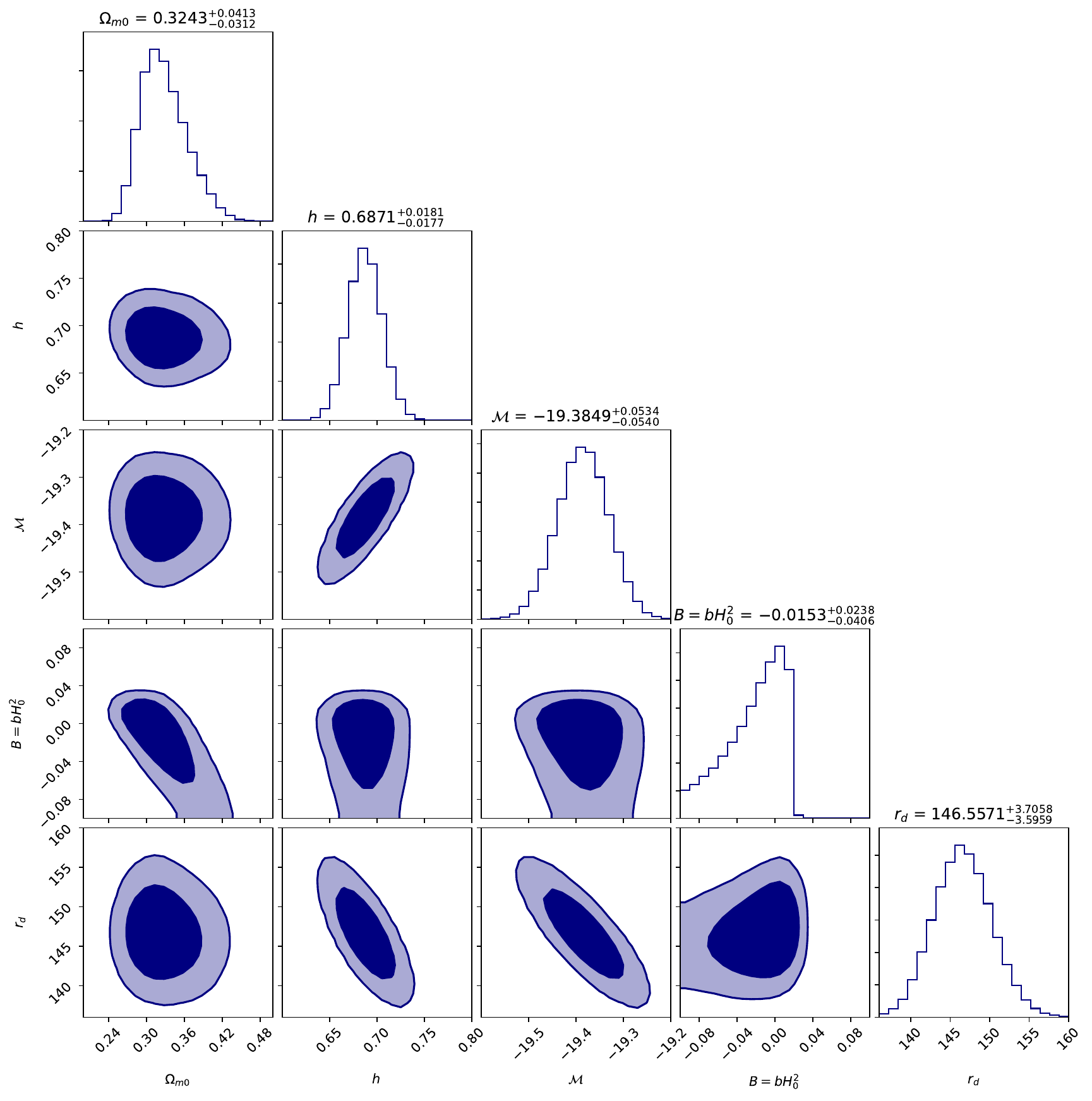}
\caption{ {\it{The $1\sigma$ and  $2\sigma$  iso-likelihood 
contours for the Quartic model (\ref{quartic1}), for the joint analysis  
SNIa+CC+BAOs, for the 2D subsets of the parameter space
$(\Omega_{m0},h,\mathcal{M},B= b H_0^2)$, using graphic package getdist 
\cite{lewis2019getdist}.  }}
}
\label{datafitquartic}
\end{figure}

We close this subsection by a detailed observational confrontation. We 
follow the steps described above, and we present the results on the parameters 
in Table \ref{tab:quartic}.  Additionally, in Fig.~\ref{datafitquartic} we 
present the corresponding contour plots 
for the inferred model parameters across all considered datasets. As a further 
consistency check, we compare the inferred value of the sound horizon at the 
baryon-drag epoch, $r_d$, with the model-independent estimate of 
\cite{Haridasu:2018gqm}, finding agreement at the $1\sigma$ level.  

As we observe, the quartic model provides a stable and well-behaved 
fit to the observational data, with all parameters being tightly constrained 
and exhibiting smooth, closed confidence regions. In particular, the matter 
density parameter $\Omega_{m0}$ and the Hubble constant $h$ are found to lie 
well within the ranges reported by recent observational analyses, indicating 
that the model successfully reproduces the main features of the late-time 
cosmic expansion.

Concerning the additional model parameter $B=bH_0^2$, we find that it is 
consistent with small but non-zero values, indicating that deviations from 
$\Lambda$CDM are allowed, although not required by the current data. 
Importantly, the contours include  $B\to 0$, which corresponds 
to the $\Lambda$CDM limit, showing that the latter is smoothly embedded within 
the present framework. This behavior is expected, since the quartic correction 
acts as a higher-curvature modification that becomes increasingly suppressed 
at low energies.

Moreover, we observe mild correlations between $B$ and the background 
parameters, particularly $\Omega_{m0}$ and $h$, reflecting the degeneracy 
between modified-gravity effects and standard cosmological components in 
driving the expansion history. Nevertheless, these degeneracies are not strong 
enough to spoil parameter determination, as evidenced by the well-localized 
posterior distributions. In summary,  the overall efficiency of the fit  
confirms that 
the quartic model 
achieves an excellent agreement with the combined SNIa, CC, and BAO datasets.

 Proceeding to a quantitative assessment, in Table~\ref{tab:AIC_BIC_quartic} 
we evaluate the performance of the quartic model against the $\Lambda$CDM 
paradigm using standard information criteria, namely the Akaike Information 
Criterion (AIC) and the Bayesian Information Criterion (BIC), following 
\cite{Anagnostopoulos:2019miu} and references therein. 
For Gaussian likelihoods, the AIC estimator is defined as
\begin{equation}
{\rm AIC}=-2\ln{\cal L}_{\rm max}+2k
=\chi^2_{\rm min}+2k ,
\end{equation} 
while the BIC is given by
\begin{equation}
\mathrm{BIC} = -2\ln \mathcal{L}_{\mathrm{max}} + k \ln N,
\end{equation}
where $\mathcal{L}_{\mathrm{max}}$ is the maximum likelihood, $N$ denotes 
the total number of data points, and $k$ is the number of free parameters 
of the model. In our analysis, we use $N=77$ data points 
($40$ SNIa, $31$ CC, and $6$ BAO measurements), while the Quartic model 
contains $k=5$ free parameters, compared to $k=4$ for $\Lambda$CDM.
In order to compare different models, we consider the relative differences 
$\Delta\mathrm{AIC} = \mathrm{AIC}_{\mathrm{model}} - \mathrm{AIC}_{\min}$ 
and similarly for $\Delta\mathrm{BIC}$. According to standard criteria, 
values $\Delta\mathrm{AIC} \le 2$ indicate statistical compatibility between 
models, while $\Delta\mathrm{AIC} \gtrsim 4$ ($\gtrsim 10$) corresponds to 
positive (strong) evidence against the model with the higher AIC value.

\begin{table}[ht]
\centering
\begin{tabular}{lccccc}
\hline\hline
Model & AIC & $\Delta$AIC & BIC & $\Delta$BIC \\
\hline
Quartic (NPQTG) & 67.097 & 1.857 & 78.816 & 4.200 \\
$\Lambda$CDM    & 65.240 & 0      & 74.615 & 0      \\
\hline\hline
\end{tabular}
\caption{Information criteria for the Quartic model (\ref{quartic1}) and 
$\Lambda$CDM, 
using the combined SNIa/CC/BAO dataset. The differences are defined as $\Delta 
\mathrm{IC} = \mathrm{IC} - \mathrm{IC}_{\min}$.}
\label{tab:AIC_BIC_quartic}
\end{table}

 As we observe, according to the AIC criterion the two scenarios are
statistically compatible and provide a comparable description of the 
data.
The BIC criterion, which penalizes more strongly the presence of additional
parameters, mildly favors the $\Lambda$CDM scenario.  Hence, the present
analysis does not provide evidence for a preference of the Quartic model over
$\Lambda$CDM. Rather, it demonstrates that the Quartic model remains allowed
by the current data and can reproduce the observed late-time expansion history
with a fitting performance comparable to that of the concordance cosmology.

In summary, this illustrative example demonstrates that the Quartic model of 
non-polynomial quasi-topological gravity    can give rise to a rich and viable 
cosmological phenomenology, compatible with current observations while allowing 
for dynamical deviations from the standard $\Lambda$CDM paradigm.
Importantly, the algebraic nature of the modified Friedmann equation allows 
for a direct and efficient comparison with observational data, making the 
framework particularly suitable for phenomenological studies.

\subsection{Power-law model }

Let us now proceed to the investigation of the Power-law model 
(\ref{powerlaw}), which gives rise to the 
effective dark energy density and pressure (\ref{rhoDE11b}),(\ref{pDE11b}), 
respectively, focusing again on the dust-matter case and neglecting radiation.  
 In this case, in contrast to the Quartic model, due to the non-trivial 
$\delta$-exponent, the scenario   does not admit a 
closed-form analytic solution for $H^2(z)$ for arbitrary $\delta$. Therefore, 
the cosmological evolution must be determined numerically by solving this 
algebraic equation at each redshift.

Once the Hubble parameter is obtained, we can directly compute the 
density parameters $\Omega_m(z)$ and $\Omega_{DE}(z)$, as well as the 
dark-energy equation-of-state parameter $w_{DE}(z)$ from (\ref{wdepower}), 
and the deceleration parameter using (\ref{deceleration}).

\begin{figure}[ht]
 \hspace{-0.7cm}
\includegraphics[scale=0.3]{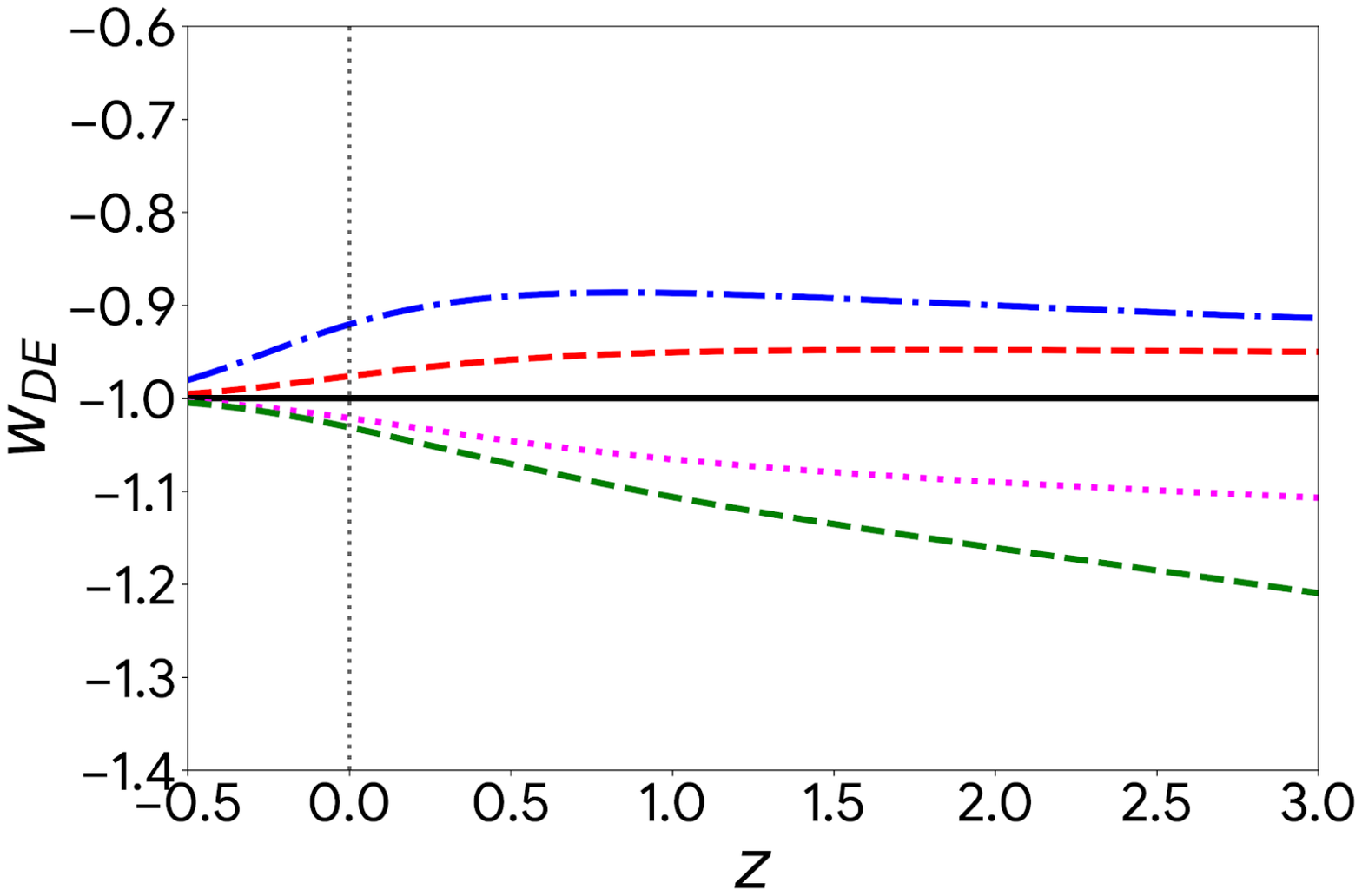}
\caption{
{\it{The evolution of the equation-of-state parameter $w_{DE}$ of   the 
Power-law 
model (\ref{powerlaw}), 
as a function of the redshift $z$,    for 
$b=1$ and for $\delta=0$   (black-solid), $\delta=-0.3$ 
(blue-dashed-dotted),  $\delta=-0.1$
(red-dashed),  $\delta=+0.1$ (pink-dotted)
 and  $\delta=+0.15$ 
(green-dashed), in units 
where $H_0=1$.
In all graphs we have imposed 
$\Omega_{DE}(z=0)\equiv\Omega_{DE0}\approx0.69$ at present,  and we 
have added a vertical dotted line denoting 
the current time $z=0$.}} }
\label{wmultiplotpower}
\end{figure}

The basic late-time cosmological evolution is qualitatively similar to
that shown in  Fig.~\ref{Omegasquartic}, exhibiting a matter-dominated epoch 
followed by
a dark-energy-dominated phase, while asymptotically the Universe
approaches complete dark-energy domination.
Additionally, the transition from decelerated to accelerated expansion happens 
at $z\approx 0.5$.  In order to examine the behavior of $w_{DE}(z)$ in more 
detail, in Fig.~\ref{wmultiplotpower} 
we display the evolution of $w_{DE}(z)$ for different choices of the 
model parameters $b$ and $\delta$. As expected, for $b=0$ the model reduces 
to $\Lambda$CDM. Increasing the magnitude of $b$ enhances the deviation 
from the standard scenario, while the parameter $\delta$ controls the 
  modification, determining how rapidly the 
correction term becomes relevant. In particular, for larger deviations of
$\delta$  from zero we obtain   larger  deviations from    $\Lambda$CDM 
scenario, as expected. Note that  $w_{DE}(z)$ can lie in the quintessence or 
phantom regime, which is an additional advantage.

\begin{table*}[!]
\centering
\begin{tabular}{lccccccc}
\hline\hline
Model & $\Omega_{m0}$ & $h$ & $r_d$ & $\delta$ & $\mathcal{M}$ & 
$\chi^2_{\min}$ 
& $\chi^2_{\min}/\mathrm{dof}$ \\
\hline
Power-law NPQTG 
& $0.3130^{+0.0345}_{-0.0211}$ 
& $0.6818^{+0.0139}_{-0.0153}$ 
& $149.2145^{+3.1812}_{-3.6414}$ 
& $0.2622^{+0.2345}_{-0.1678}$ 
& $-19.4083^{+0.0410}_{-0.0497}$ 
& $56.842$ 
& $0.790$ \\
\hline
\end{tabular}
\caption{Observational constraints for the Power-law   model (\ref{powerlaw}) 
(with $b=1$) 
from the joint SNIa/CC/BAO analysis.}
\label{tab:powerlaw}
\end{table*}

\begin{figure}[!]
\includegraphics[width=0.48\textwidth]{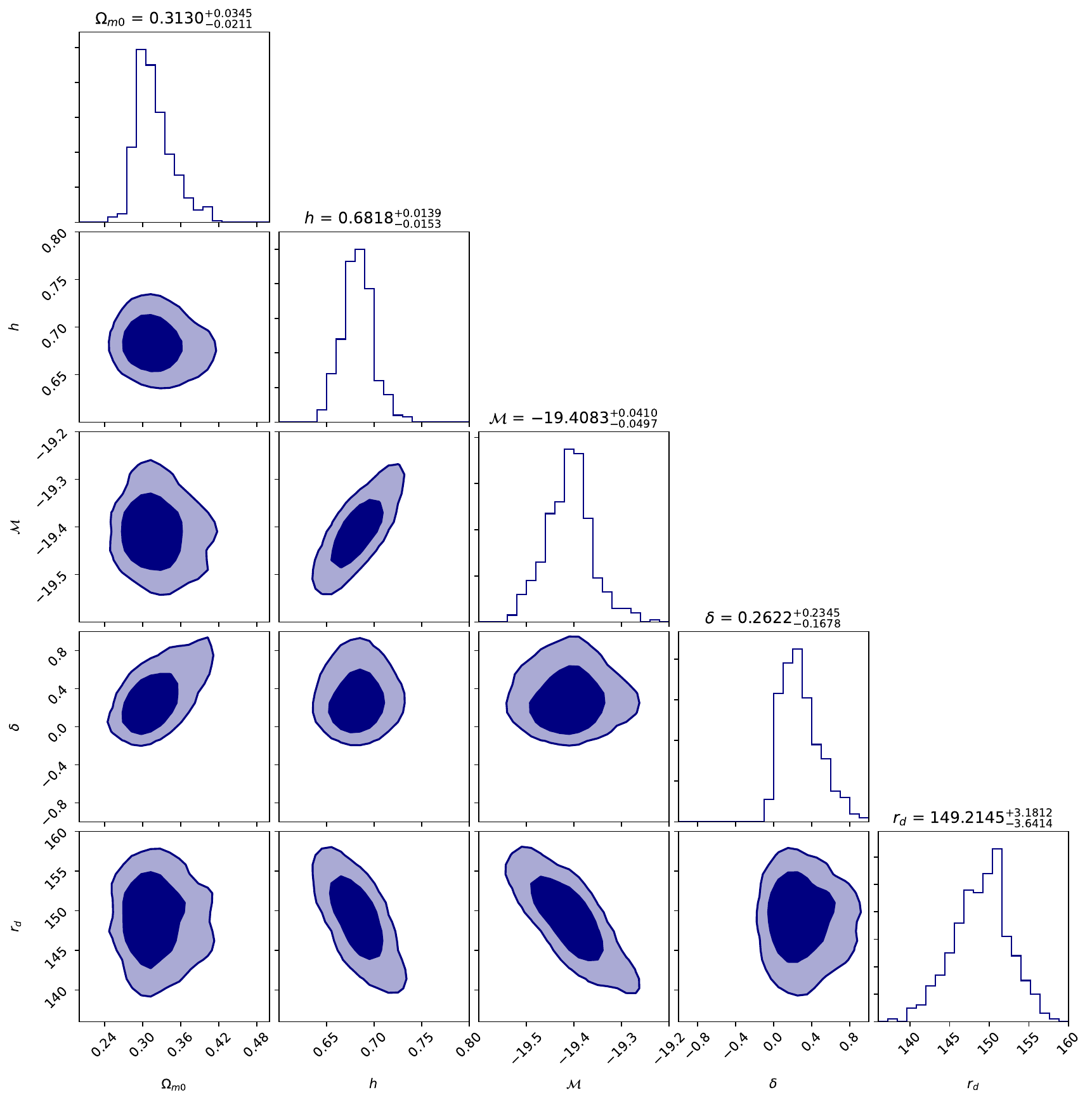}
\caption{ {\it{The $1\sigma$ and  $2\sigma$  iso-likelihood 
contours for the Power-law model (\ref{powerlaw}), for the joint analysis  
SNIa+CC+BAOs, for the 2D subsets of the parameter space
$(\Omega_{m0},h,\mathcal{M},\delta)$ using graphic 
package getdist 
\cite{lewis2019getdist}.  }}
}
\label{Datafitpower}
\end{figure}

We now proceed to the observational analysis. Following the same 
methodology as described previously, we perform a joint fit to the 
SNIa+CC+BAOs datasets, and we summarize the resulting parameter 
constraints in Table~\ref{tab:powerlaw}. Furthermore, in 
Fig.~\ref{Datafitpower} 
we present the corresponding contour plots for the parameter space. We mention 
here that since the simultaneous variation of $b$ and $\delta$
 leads to a strong normalization degeneracy in the background dynamics, we fix 
$b=1$ in order to isolate the effect of the exponent  $\delta$.

As we see,   the standard cosmological 
parameters, such as $\Omega_{m0}$ and $h$, are constrained within ranges fully 
compatible with independent measurements, indicating that the model 
successfully reproduces the observed expansion history. 
Focusing on the additional parameter $\delta$, which governs the deviation from 
$\Lambda$CDM, we observe that its posterior distribution is centered close to 
zero, with relatively broad but well-defined confidence intervals. This 
behavior implies that the $\Lambda$CDM scenario is naturally recovered within 
the parameter space of the model, while at the same time allowing for mild 
dynamical departures that remain consistent with current data.  
Moreover, the contour structure indicates moderate degeneracies between 
$\delta$ and the background parameters, especially $\Omega_{m0}$, reflecting 
the interplay between modified-gravity effects and the matter sector in shaping 
the cosmic evolution. Nevertheless, these degeneracies remain sufficiently 
controlled, ensuring a stable parameter estimation and a well-defined 
likelihood landscape.

Finally, we assess the statistical performance of the Power-law model 
against $\Lambda$CDM using AIC and BIC information criteria, presenting the 
results 
in Table \ref{tab:AIC_BIC_powerlaw}. 
As we observe,  the differences in the information criteria remain small,
showing that the Power-law model provides a quality of fit comparable to that
of $\Lambda$CDM. Nevertheless, the BIC mildly favors the latter due to its
smaller parameter space. Therefore, the present data do not indicate a
statistically significant preference for the Power-law modification, but they
show that it remains a viable cosmological scenario compatible with the
observed late-time expansion history.

\begin{table}[ht]
\centering
\begin{tabular}{lcccc}
\hline\hline
Model & AIC & $\Delta$AIC & BIC & $\Delta$BIC \\
\hline
Power-law (NPQTG) & 66.842 & 1.541 & 78.561 & 3.885 \\
$\Lambda$CDM & 65.301 & 0 & 74.676 & 0 \\
\hline
\end{tabular}
\caption{Information criteria for the  Power-law   model (\ref{powerlaw})  
(with 
$b=1$) and 
$\Lambda$CDM, using the combined SNIa/CC/BAO dataset. The differences are 
defined as $\Delta$IC $=$ IC $-$ IC$_{\min}$.}
\label{tab:AIC_BIC_powerlaw}
\end{table}

In summary, the Power-law model constitutes a natural generalization 
of the Quartic scenario, offering additional flexibility through the 
parameter $\delta$, while preserving the algebraic simplicity of the 
framework. This allows for a rich phenomenology, capable of describing 
late-time cosmic acceleration with dynamical dark energy, without 
departing significantly from observational constraints.

\section{Conclusions}

In this work we investigated the cosmological implications of  a
phenomenologically relevant subclass  of non-polynomial quasi-topological
gravity, for which the background dynamics can be encoded in a single
function $h(H^2)$. This framework is particularly appealing, 
since it allows for the incorporation of higher-curvature corrections 
while preserving a remarkably simple cosmological structure, avoiding 
the higher-order differential equations that typically arise in 
generalized gravity theories. In this sense, non-polynomial 
quasi-topological gravity provides a powerful and flexible setting for 
constructing viable extensions of general relativity, capable of 
capturing deviations at high curvature while maintaining consistency 
with the well-tested low-energy regime.

Within this general framework, we first established the conditions for 
cosmological viability and demonstrated that a wide class of modified 
Friedmann equations can be consistently realized from an underlying 
action. In particular, we constructed both polynomial and genuinely 
non-polynomial realizations, explicitly deriving the corresponding 
reduced action functions. These constructions reveal  the capabilities 
of the theory, showing that a broad family of cosmological models, ranging 
from simple power-law corrections to more intricate non-polynomial 
structures, can be embedded in a unified geometric description.

Focusing on concrete realizations, we analyzed in detail the Quartic
and Power-law models, which constitute representative and
phenomenologically relevant cases. In both scenarios,  within the
late-time regime investigated here, the cosmological evolution exhibits
a matter-dominated epoch followed by a transition to an accelerated
phase. Importantly, the modified gravitational sector gives rise to an
effective dark-energy component with dynamical properties.  In particular, the 
corresponding equation-of-state parameter 
can evolve in time and, in specific regions of the parameter space, enter 
the phantom regime without introducing pathological degrees of freedom. 
This feature constitutes a significant advantage of the present 
framework, as it allows for rich phenomenology beyond the $\Lambda$CDM 
paradigm while remaining theoretically well-defined.  We emphasize that the
specific quartic realization   is
regarded as an effective late-time model and is not extrapolated to the
radiation-dominated or arbitrarily high-curvature regimes.

The observational confrontation of the models with the SNIa, CC, and BAO
datasets further supports their cosmological viability.  For the Quartic 
model,
the parameter controlling the higher-curvature correction remains compatible
with the $\Lambda$CDM limit, while allowing for controlled departures from the
standard scenario. Similarly, in the Power-law case, the exponent parameter
permits mild dynamical deviations while remaining consistent with the current
constraints. In both cases, the models reproduce the observed late-time
expansion history with a quality of fit comparable to that of $\Lambda$CDM.
The information-criteria analysis leads to the same conclusion, namely  the AIC 
shows
statistical compatibility between the scenarios, whereas the BIC mildly favors
$\Lambda$CDM because of its smaller parameter space. Hence, the present data do
not provide evidence for a departure from the concordance cosmology, but rather 
they demonstrate that the NPQTG models considered here remain viable
alternatives capable of accommodating dynamical dark-energy behavior.

The present analysis opens several promising directions for future 
investigation. A natural extension would involve the study of 
cosmological perturbations, in order to examine the growth of structures 
and confront the models with large-scale structure and CMB data. 
Additionally, the exploration of early-Universe implications, such as 
inflationary dynamics or the avoidance of cosmological singularities, 
could reveal further distinctive signatures of the theory. Finally, the 
systematic classification of non-polynomial functions $h(H^2)$ and their 
connection to underlying fundamental theories may provide deeper insight 
into the geometric origin of cosmic acceleration. We therefore believe 
that non-polynomial quasi-topological gravity constitutes a compelling 
framework for extending our understanding of gravitational physics and 
cosmology beyond the standard paradigm.

 \begin{acknowledgments}
  The author acknowledges the contribution of the LISA   CosWG  and of   COST
Actions  
 CA21106 ``COSMIC WISPers
in the Dark Universe: Theory, astrophysics and experiments'',  CA21136 
``Addressing observational tensions in cosmology with 
  systematics and fundamental physics (CosmoVerse)'',    CA23130 
``Bridging high and low energies in
search of quantum gravity (BridgeQG)'', and CA24101 ``Testing Fundamental 
Physics with Seismology''.

 \end{acknowledgments}
 
 
\end{document}